\documentclass[%
  twocolumn,
 showpacs,
 showkeys,
 preprintnumbers,
 amsmath,amssymb,
 aps,
  pra,
  longbibliography,
 floatfix,
 ]{revtex4-1}

\usepackage{mathptmx}% http://ctan.org/pkg/mathptmx

\usepackage{amssymb,amsthm,amsmath}

\usepackage{tikz}
\usepackage[breaklinks=true,colorlinks=true,anchorcolor=blue,citecolor=blue,filecolor=blue,menucolor=blue,pagecolor=blue,urlcolor=blue,linkcolor=blue]{hyperref}
\usepackage{graphicx}% Include figure files
\usepackage{url}

\usepackage{xcolor}

\begin{document}

\title{Entanglement through path identification}

\author{Karl Svozil}
\email{svozil@tuwien.ac.at}
\homepage{http://tph.tuwien.ac.at/~svozil}

\affiliation{Institute for Theoretical Physics,
Vienna  University of Technology,
Wiedner Hauptstrasse 8-10/136,
1040 Vienna,  Austria}

\affiliation{Department of Computer Science, University of Auckland,
Private Bag 92019, Auckland, New Zealand}

\date{\today}

\begin{abstract}
Entanglement in multipartite systems can be achieved by the coherent superposition of product states, generated through a universal unitary transformation, followed by spontaneous parametric down-conversions and path identification.
\end{abstract}

%\pacs{03.65.Aa, 03.65.Ta, 03.65.Ud, 03.67.-a}
%\preprint{CDMTCS preprint nr. x}

\maketitle

From a formal point of view, an arbitrary pure (we shall not consider mixed states as we consider them epistemic)
state of $N$ particles with dichotomic properties $0$, $1$
can be written as the coherent superposition
\begin{equation}
\begin{split}
 \vert   \Psi   \rangle
=
\sum_{i_1,\ldots , i_N=0}^{1}
\alpha_{i_1,\ldots , i_N}
\vert   i_1,\ldots , i_N    \rangle
\text{ with } \\
\sum_{i_1,\ldots , i_N=0}^{1}
\vert \alpha_{i_1,\ldots , i_N} \vert^2 =   1
\end{split}
\label{2017-etpi-e1}
\end{equation}
of all product states
$\vert   i_1,\ldots , i_N    \rangle
=
\vert    i_1 \rangle  \cdots  \vert   i_N    \rangle
$.
One possible direct physical implementation of this formula
requires
(i) a universal (with respect to the unitary group) transformation
rendering the coefficients
$\alpha_{i_1,\ldots , i_N}$;
followed by
(ii)   a  spontaneous parametric down-conversion producing the product states
whose outputs are properly integrated and identified in a third phase (iii).

In what follows we shall use Fock states (notwithstanding issues such as localization~\cite[p.~931]{mandel-PhysRevA.28.929})
having definite occupation numbers of the quantized field modes.
For such states the unitary quantum evolution on elementary quantum optical components
can be represented by elementary transition rules, reflecting unitary transformations~\cite{zeilinger:882,green-horn-zei}:
a symmetrical beam splitter is represented by
$
\vert \text{in} \rangle
\xrightarrow{50:50\; \text{BS}}
\frac{1}{\sqrt{2}}\left(
\vert \text{transit} \rangle
+ i
\vert \text{reflect} \rangle
\right)
$; and an asymmetrical beam splitter by
$
\vert \text{in} \rangle
\xrightarrow{\text{BS}}
T
\vert \text{transit} \rangle
+ i
R
\vert \text{reflect} \rangle
$, with
$\vert T \vert^2
+
\vert R \vert^2 =1$.
Phase shift(er)s are represented by
$
\vert \text{in} \rangle
\xrightarrow{\varphi \;\text{ps}}
e^{i\varphi}
\vert \text{in} \rangle
$,
and  spontaneous parametric down-conversions by
$
\vert \text{in} \rangle
\xrightarrow{\text{NL}}
\eta \vert \text{out}1 \rangle \vert  \text{out}2 \rangle
=
\eta \vert \text{out}1\, \text{out}2 \rangle
$  for supposedly small $\eta$.

\begin{figure}
\begin{center}
\includegraphics[width=8cm,angle=0]{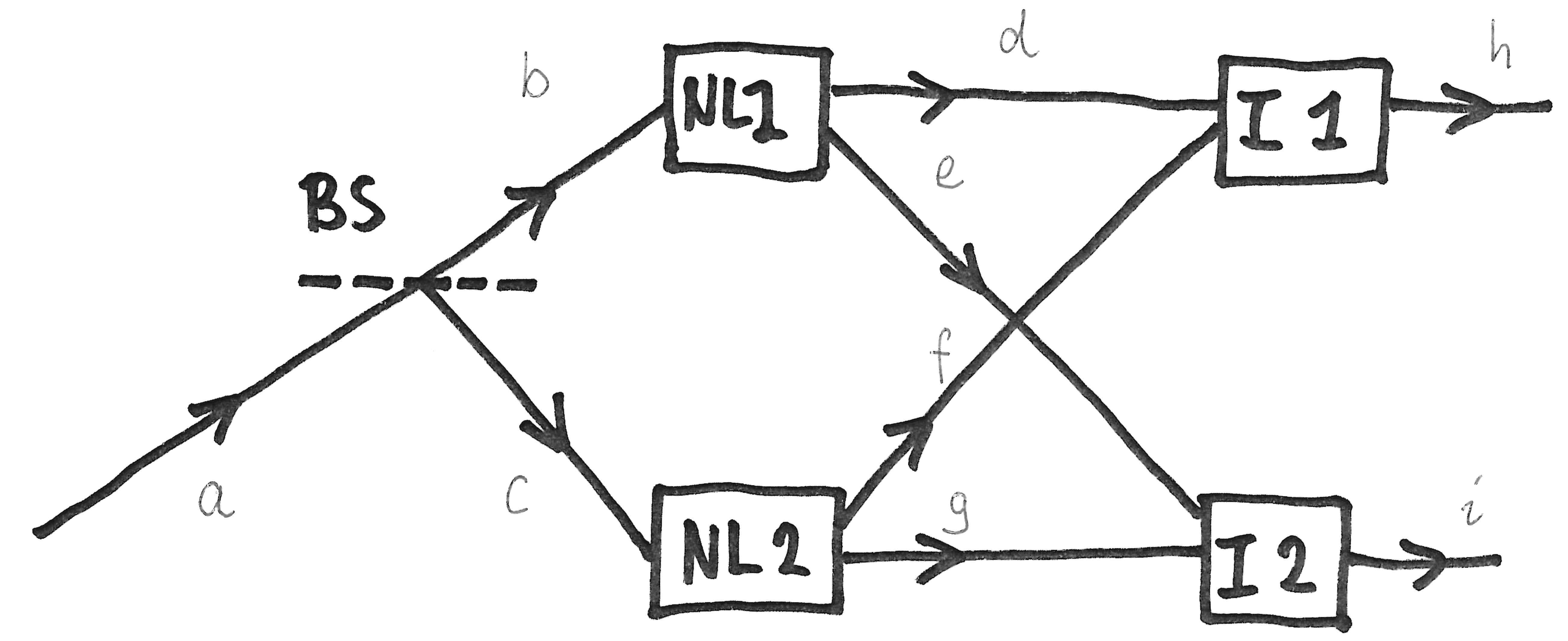}
\end{center}
\caption{
An interferometric experiment involving an incident beam, a beam splitter, and two spontaneous parametric down-conversion crystals.
\label{2017-etpi-f1}
}
\end{figure}

For the sake of a demonstration, consider an arrangement depicted in Fig.~\ref{2017-etpi-f1}.
It consists of a single particle source producing a state
$\vert a \rangle$
impinging on a symmetrical beam splitter
$\text{BS}$
whose output ports are identified with the states
$\vert b \rangle$  for transiting $\vert a \rangle$, and
$\vert c \rangle$  for reflected $\vert a \rangle$, respectively.
Those states are the subjected to  two
spontaneous parametric down-conversion crystals
$\text{NL1}$ and
$\text{NL2}$, producing product pairs
$\vert d  e \rangle$ and
$\vert f  g \rangle$, respectively.
``Adjacent'' beam pairs $d$--$f$ as well as $e$--$g$
are then integrated and identified a states
$\vert h \rangle$ and
$\vert i \rangle$,
respectively.
The aforementioned substitution rules
yield
\begin{equation}
\begin{split}
 \vert   a   \rangle
\xrightarrow{50:50\; \text{BS}}
\frac{1}{\sqrt{2}}
\left(
\vert b \rangle
+ i
\vert c \rangle
\right)
\xrightarrow{\text{NL1, NL2}}
\frac{\eta }{\sqrt{2}}
\left(
\vert de \rangle
+ i
\vert fg \rangle
\right).
\end{split}
\label{2017-etpi-e2}
\end{equation}
Note that an additional phase shift of $\varphi =\frac{\pi}{2}$ applied to $\vert c \rangle$,
with the identification $d=g=0$ and $e=f=1$,
would have resulted in the traditional singlet state
$\vert \Psi^- \rangle  =
\frac{1}{\sqrt{2}}
\left(
\vert 01 \rangle
-
\vert 10 \rangle
\right)
$ of the Bell basis.

The final phase of this experiment is depicted in Fig.~\ref{2017-etpi-f1} by the addition of ``integrators''
$\text{I1}$
and
$\text{I2}$
which combine or collimate ingoing ports into a single port.
Thereby it is not necessary to take care that it is impossible for any observer to determine from which
of the spontaneous parametric down-conversion crystals the quantum came from.
The which-way information may be obtained through measurement of
the output
$
\vert h \rangle
$
or
$
\vert i \rangle
$.

In order to fully realize Eq.~(\ref{2017-etpi-e1}),
universal unitary transformations in finite-dimensional Hilbert space
need to be operationalized.
One conceivable way of doing this is through generalized beam splitters~\cite{rzbb},
which is based upon the parameterization of the unitary group~\cite{murnaghan}.
Fig.~\ref{2017-etpi-f1} depicts this configuration for two dichotomic (two possible states per quantum) quanta.
A generalization to an arbitrary number of quanta, as well as an arbitrary number of states per quanta
can be given along very similar lines.

\begin{figure}
\begin{center}
\includegraphics[width=8cm,angle=0]{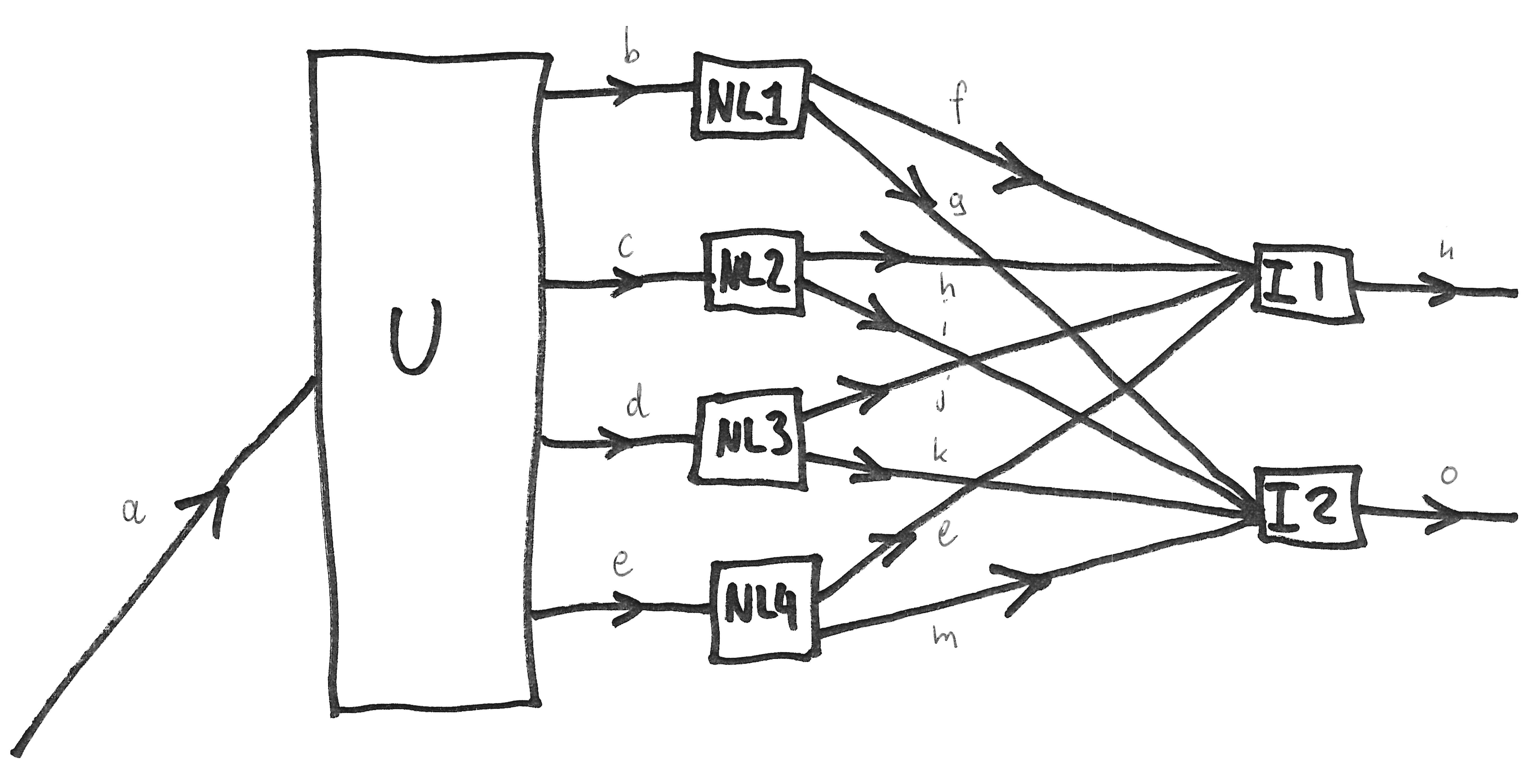}
\end{center}
\caption{
General two-particle state generation.
\label{2017-etpi-f2}
}
\end{figure}

Let me finally address the question why, even granted the fact that this might be a novel way of looking at and producing multipartite states
(I am quite confident that similar schemes might have been proposed in one way or another before, but I am unaware and thus less than sure about these),
one needs yet another scheme.
After all, higher-dimensional two-particle entanglements can be realized in principle solely via multiport beam splitters~\cite{zukowski-97};
without some additional final steps involving
spontaneous parametric down-conversion and integration.
(This conforms to the interpretation of the Clauser-Horne-Shimony-Holt expression as a single operator which can be
subjected to min-max considerations~\cite{filipp-svo-04-qpoly-prl}.)
It should also be mentioned that a recent proposal~\cite{PhysRevLett.118.080401}, based on an intriguing experiment~\cite{zou-wang-mandel:91a,zou-wang-mandel:91b}
upon a suggestion of Ou~\cite{ou-2007}, uses path identification as a resource to produce multipartite states.

One good motivation for the aforementioned contemplations might be the ``production'' of entanglement in these configurations
which might yield fresh ways to perceive or ``understand'' this quantum feature.
As expressed by Bennett~\cite{Bennett-IBM-03.05.2016}
in quantum physics the possibility exists {\em ``that you have a complete knowledge
of the whole without knowing the state of any one part. That a thing can be in a definite
state, even though its parts were not.~$\ldots$
It's not a complicated idea but
it's an idea that nobody would ever think of.''}
Bennett, if I interpret him correctly, is referring to
Schr\"odingers's 1935 \& 1936 series of papers; both in German~\cite{schrodinger,schrodinger-en-10.2307/986572}
and English~\cite{CambridgeJournals:1737068,CambridgeJournals:2027212}
pointing out that the quantum state of multiple particles can evolve in such ways that,
say,
the initial definiteness of the states of the individual independent constituent without any relational properties among themselves
gets re-encoded into purely relational properties among the particles~\cite{wootters-1990-localaccOQStates,mermin:753,Zeilinger-97,zeil-99},
thereby ``erasing'' the definiteness of the individual particle properties.
One may also say that the multipartite state is ``breathing in and out of'' individuality and entanglement~\cite{svozil-2016-pu-book}.

The formal expression for this is a sort of zero-sum game with respect to knowledge or information encoded by the quantum state:
due to the permutative character of the unitary (one-to-one isometry) state evolution, no information is ever lost or gained;
that is, any loss of individual definiteness ``on'' the individual constituents has to be compensated by a gain through
``sampling'' of their independence; to the effect that they are no longer independent but possess definite relational properties.
Conversely, any ``scrambling'' of these relational properties needs to be (due to the impossibility to ``loose'' information)
compensated by a gain of individual definitiveness.

For the sake of a concrete demonstration~\cite[Section~1.5]{mermin-07},
consider a a general state in 4-dimensional Hilbert space.
It can be written as a vector in ${\Bbb C}^4$, which can be parameterized by
\begin{equation}
\begin{pmatrix}\alpha_1,\alpha_2,\alpha_3,\alpha_4\end{pmatrix}^\intercal , \textrm{ with } \alpha_1,\alpha_3,\alpha_3,\alpha_4 \in {\Bbb C},
\label{2012-m-ch-fdvs-dectp-gf}
\end{equation}
and suppose (wrongly)  (\ref{2012-m-ch-fdvs-dectp-gf}) that all such states can be written in terms of a tensor product of two quasi-vectors in  ${\Bbb C}^2$
\begin{equation}
\begin{pmatrix}a_1,a_2\end{pmatrix}^\intercal \otimes \begin{pmatrix}b_1,b_2\end{pmatrix}^\intercal
\equiv \begin{pmatrix}a_1b_1, a_1 b_2,a_2b_1,a_2b_2\end{pmatrix}^\intercal ,
\label{2012-m-ch-fdvs-dectp-gftp}
\end{equation}
with $a_1,a_2,b_1,b_2\in {\Bbb C}$.
A comparison of the coordinates in
(\ref{2012-m-ch-fdvs-dectp-gf})
and
(\ref{2012-m-ch-fdvs-dectp-gftp})
yields
\begin{equation}
\begin{split}
\alpha_1=a_1b_1,\quad
\alpha_2=a_1b_2,\quad
\alpha_3=a_2b_1,\quad
\alpha_4=a_2b_2.
\end{split}
\label{2012-m-ch-fdvs-dectp-gftp-a}
\end{equation}
By taking the quotient of the two first and the two last equations, and by equating these quotients, one obtains
\begin{equation}
\begin{split}
\frac{\alpha_1}{\alpha_2}=\frac{b_1}{b_2}
=\frac{\alpha_3}{\alpha_4},\textrm{ and thus }
{\alpha_1}{\alpha_4}={\alpha_2}{\alpha_3}.
\end{split}
\label{2012-m-ch-fdvs-dectp-gftp-fr}
\end{equation}

How can we imagine this?
As in many cases, states in the Bell basis, and, in particular, the Bell state,
serve as a sort of
{\em Rosetta Stone}
\index{Rosetta Stone}
for an understanding of this quantum feature.
The {\em Bell state}
\index{Bell state}
$\vert \Psi^- \rangle$ is a typical example of an entangled state;
or, more generally, states in the
{\em Bell basis}
\index{Bell basis} can be defined and,
with
$\vert  0 \rangle =  \begin{pmatrix}1,0\end{pmatrix}^\intercal$
and
$\vert  1 \rangle =  \begin{pmatrix}0,1\end{pmatrix}^\intercal$
  encoded by
\begin{equation}
\begin{split}
\vert \Psi^\mp \rangle = \frac{1}{\sqrt{2}}\left(\vert 0   1 \rangle \mp \vert 1   0 \rangle  \right)= \begin{pmatrix} 0, 1, \mp 1, 0\end{pmatrix}^\intercal ,  \\
\vert \Phi^\mp \rangle = \frac{1}{\sqrt{2}}\left(\vert 0   0 \rangle \mp \vert 1   1 \rangle  \right)= \begin{pmatrix} 1, 0, 0, \mp 1\end{pmatrix}^\intercal  .
\end{split}
\label{2014-m-ch-fdvs-bellbasis2}
\end{equation}
For instance, in the case of $\vert \Psi^- \rangle$ a comparison of coefficient yields
\begin{equation}
\begin{split}
\alpha_1=a_1b_1=0, \quad
\alpha_2=a_1b_2=\frac{1}{\sqrt{2}},\\
\alpha_3=a_2b_1-\frac{1}{\sqrt{2}}, \quad
\alpha_4=a_2b_2=0;
\end{split}
\label{2012-m-ch-fdvs-BellSCC}
\end{equation}
and thus
entanglement, since
\begin{equation}
{\alpha_1}{\alpha_4}=0 \neq {\alpha_2}{\alpha_3}=\frac{1}{2}.
\end{equation}
This shows that  $\vert \Psi^- \rangle$ cannot be considered as a two particle product state.
Indeed, the state can only be characterized by considering the {\em relative properties}
of the two particles --
in the case of  $\vert \Psi^- \rangle$ they are associated with the statements~\cite{zeil-99}:
``the quantum numbers (in this case ``$0$'' and ``$1$'') of the two particles are different in (at least) two orthogonal directions.''

The Bell basis symbolizing entanglement and nonindividuality can, in an {\it ad hoc} manner, be generated from a nonentangled, individual state:
suppose such a styte is represented by elements of the Cartesian standard basis
in $4$-dimensional real space ${\Bbb R}^4$, representable as column vectors whose components are
$\begin{pmatrix} \vert {\bf e}_i \rangle \end{pmatrix}_j = \delta_{ij}$, with $1\le i,j \le 4$.
%\begin{equation}
%\begin{split}
%\vert {\bf e}_1 \rangle = \begin{pmatrix}1\\ 0\\ 0\\ 0\end{pmatrix},\,
%\vert {\bf e}_2 \rangle = \begin{pmatrix}0\\ 1\\ 0\\ 0\end{pmatrix},\,
%\vert {\bf e}_3 \rangle = \begin{pmatrix}0\\ 0\\ 1\\ 0\end{pmatrix},\,
%\vert {\bf e}_4 \rangle = \begin{pmatrix}0\\ 0\\ 0\\ 1\end{pmatrix}.
%\end{split}
%\label{2016-m-fdvs-csb4}
%\end{equation}
Suppose further that the coordinates
of the Bell basis~(\ref{2014-m-ch-fdvs-bellbasis2}) are arranged as row or column vectors, thereby forming the respective unitary transformation
\begin{equation}
\begin{split}
\textsf{\textbf{U}} =
\vert \Psi^- \rangle \langle {\bf e}_1  \vert  +
\vert \Psi^+ \rangle \langle {\bf e}_2  \vert  +
\vert \Phi^- \rangle \langle {\bf e}_3  \vert  +
\vert \Phi^+ \rangle \langle {\bf e}_4  \vert
=
\\
=
\begin{pmatrix}
\vert \Psi^- \rangle ,
\vert \Psi^+ \rangle ,
\vert \Phi^- \rangle ,
\vert \Phi^+ \rangle  \end{pmatrix}
=   \frac{1}{\sqrt{2}}
\begin{pmatrix}
0 & 0 &   1 &   1 \\
1 & 1 &   0 &   0 \\
-1& 1 &   0 &   0 \\
0 & 0 &  -1 &   1
 \end{pmatrix}
.
\end{split}
\label{2014-m-ch-fdvs-bellbasis2m}
\end{equation}
% Eigensystem[(1/Sqrt[2]) {{0, 0, 1, 1}, {1, 1, 0, 0}, {-1, 1, 0, 0}, {0, 0,1,  -1}}]
Then successive application of  $\textsf{\textbf{U}}$ and its inverse $\textsf{\textbf{U}}^\intercal$ transforms an individual,
nonentangled state from the Cartesian basis back and forth
into an entangled, nonindividual state from the Bell basis.
For the sake of another demonstration,
consider the following perfectly cyclic evolution  which permutes all (non)entangled  states
corresponding to the Cartesian \&  Bell bases:
\begin{equation}
\begin{split}
\vert {\bf e}_1 \rangle
\stackrel{\textsf{\textbf{U}}}{\mapsto}
\vert \Psi^- \rangle
\stackrel{\textsf{\textbf{V}}}{\mapsto}
\vert {\bf e}_2 \rangle
\stackrel{\textsf{\textbf{U}}}{\mapsto}
\vert \Psi^+ \rangle   \\
\stackrel{\textsf{\textbf{V}}}{\mapsto}
\vert {\bf e}_3 \rangle
\stackrel{\textsf{\textbf{U}}}{\mapsto}
\vert \Phi^- \rangle
\stackrel{\textsf{\textbf{V}}}{\mapsto}
\vert {\bf e}_4 \rangle
\stackrel{\textsf{\textbf{U}}}{\mapsto}
\vert \Phi^+ \rangle
\stackrel{\textsf{\textbf{V}}}{\mapsto}
\vert {\bf e}_1 \rangle
.
\end{split}
\label{2017-m-pu-book-chapter-qm-baf}
\end{equation}
This evolution is facilitated by $\textsf{\textbf{U}}$
of Eq.~(\ref{2014-m-ch-fdvs-bellbasis2m}), as well as by the following additional unitary transformation~\cite{Schwinger.60}:
\begin{equation}
 \begin{split}
%\textsf{\textbf{U}} =
%\vert \Psi^- \rangle \langle {\bf e}_1  \vert  +
%\vert \Psi^+ \rangle \langle {\bf e}_2  \vert  +
%\vert \Phi^- \rangle \langle {\bf e}_3  \vert  +
%\vert \Phi^+ \rangle \langle {\bf e}_4  \vert  ,\\
\textsf{\textbf{V}} =
\vert {\bf e}_2 \rangle \langle  \Psi^-  \vert  +
\vert {\bf e}_3 \rangle \langle  \Psi^+  \vert  +
\vert {\bf e}_4 \rangle \langle  \Phi^-  \vert  +
\vert {\bf e}_1 \rangle \langle  \Phi^+  \vert
=
\\
=
\begin{pmatrix}
 \langle \Phi^+ \vert \\
 \langle \Psi^- \vert \\
 \langle \Psi^+ \vert \\
 \langle \Phi^- \vert  \end{pmatrix}
=   \frac{1}{\sqrt{2}}
\begin{pmatrix}
1 & 0 &   0 &   1 \\
0 & 1 &   -1 &   0 \\
0& 1 &   1 &   0 \\
1 & 0 &  0 &  -1
 \end{pmatrix}
.
\end{split}
\label{2017-m-pu-book-chapter-qm-baf-uo}
\end{equation}

% Eigensystem[(1/Sqrt[2]) {{1, 0, 0, 1}, {0, 1, -1, 0}, {0, 1, 1,  0}, {1, 0, 0, -1}}]

One of the ways thinking of this kind of  breathing in and out of individuality {\&} entanglement  is in terms of
sampling  {\&} scrambling of information,
as quoted from Chiao~\cite[p.~27]{green-horn-zei} (reprinted in~\cite{Macchiavello-2001}):
{\em ``Nothing has really been erased here, only scrambled!''}
Indeed, as noted earlier, mere re-coding or ``scrambling,''
and not erasure or creation of information, is tantamount to,
and an expression and direct consequence of,
the unitary evolution of the quantum state.

Let us now reconsider the configuration depicted in Fig.~\ref{2017-etpi-f1}: it is quite obvious where the relational properties
in the resulting entangled (with a proper identification)
state~(\ref{2017-etpi-e2}) come from: they reside in the common origin of either the states
$\vert d \rangle${\&}$\vert e \rangle$,
(exclusive) or
$\vert f \rangle${\&}$\vert g \rangle$,
respectively;
and in their coherent superposition rendered by the beam splitter $\text{BS}$.
This letter beam splitter $\text{BS}$ element ``scrambles'' all individuality (with respect to ``which way'' information about the output ports);
whereas the pair production at the two spontaneous parametric down-conversion crystals
is responsible for the relational -- that is, joint --  occurrence among the constituents.

\begin{acknowledgments}
This work was supported in part by  the John Templeton Foundation's {\em  Randomness and Providence: an Abrahamic Inquiry Project}.
I thank Johann Summhammer for useful comments and suggestions.

\end{acknowledgments}

%\bibliography{svozil}

\begin{thebibliography}{24}%
\makeatletter
\providecommand \@ifxundefined [1]{%
 \@ifx{#1\undefined}
}%
\providecommand \@ifnum [1]{%
 \ifnum #1\expandafter \@firstoftwo
 \else \expandafter \@secondoftwo
 \fi
}%
\providecommand \@ifx [1]{%
 \ifx #1\expandafter \@firstoftwo
 \else \expandafter \@secondoftwo
 \fi
}%
\providecommand \natexlab [1]{#1}%
\providecommand \enquote  [1]{``#1''}%
\providecommand \bibnamefont  [1]{#1}%
\providecommand \bibfnamefont [1]{#1}%
\providecommand \citenamefont [1]{#1}%
\providecommand \href@noop [0]{\@secondoftwo}%
\providecommand \href [0]{\begingroup \@sanitize@url \@href}%
\providecommand \@href[1]{\@@startlink{#1}\@@href}%
\providecommand \@@href[1]{\endgroup#1\@@endlink}%
\providecommand \@sanitize@url [0]{\catcode `\\12\catcode `\$12\catcode
  `\&12\catcode `\#12\catcode `\^12\catcode `\_12\catcode `\%12\relax}%
\providecommand \@@startlink[1]{}%
\providecommand \@@endlink[0]{}%
\providecommand \url  [0]{\begingroup\@sanitize@url \@url }%
\providecommand \@url [1]{\endgroup\@href {#1}{\urlprefix }}%
\providecommand \urlprefix  [0]{URL }%
\providecommand \Eprint [0]{\href }%
\providecommand \doibase [0]{http://dx.doi.org/}%
\providecommand \selectlanguage [0]{\@gobble}%
\providecommand \bibinfo  [0]{\@secondoftwo}%
\providecommand \bibfield  [0]{\@secondoftwo}%
\providecommand \translation [1]{[#1]}%
\providecommand \BibitemOpen [0]{}%
\providecommand \bibitemStop [0]{}%
\providecommand \bibitemNoStop [0]{.\EOS\space}%
\providecommand \EOS [0]{\spacefactor3000\relax}%
\providecommand \BibitemShut  [1]{\csname bibitem#1\endcsname}%
\let\auto@bib@innerbib\@empty
%</preamble>
\bibitem [{\citenamefont {Mandel}(1983)}]{mandel-PhysRevA.28.929}%
  \BibitemOpen
  \bibfield  {author} {\bibinfo {author} {\bibfnamefont {L.}~\bibnamefont
  {Mandel}},\ }\bibfield  {title} {\enquote {\bibinfo {title} {Photon
  interference and correlation effects produced by independent quantum
  sources},}\ }\href {\doibase 10.1103/PhysRevA.28.929} {\bibfield  {journal}
  {\bibinfo  {journal} {Physical Review A}\ }\textbf {\bibinfo {volume} {28}},\
  \bibinfo {pages} {929--943} (\bibinfo {year} {1983})}\BibitemShut {NoStop}%
\bibitem [{\citenamefont {Zeilinger}(1981)}]{zeilinger:882}%
  \BibitemOpen
  \bibfield  {author} {\bibinfo {author} {\bibfnamefont {Anton}\ \bibnamefont
  {Zeilinger}},\ }\bibfield  {title} {\enquote {\bibinfo {title} {General
  properties of lossless beam splitters in interferometry},}\ }\href {\doibase
  10.1119/1.12387} {\bibfield  {journal} {\bibinfo  {journal} {American Journal
  of Physics}\ }\textbf {\bibinfo {volume} {49}},\ \bibinfo {pages} {882--883}
  (\bibinfo {year} {1981})}\BibitemShut {NoStop}%
\bibitem [{\citenamefont {Greenberger}\ \emph {et~al.}(1993)\citenamefont
  {Greenberger}, \citenamefont {Horne},\ and\ \citenamefont
  {Zeilinger}}]{green-horn-zei}%
  \BibitemOpen
  \bibfield  {author} {\bibinfo {author} {\bibfnamefont {Daniel~M.}\
  \bibnamefont {Greenberger}}, \bibinfo {author} {\bibfnamefont {Mike~A.}\
  \bibnamefont {Horne}}, \ and\ \bibinfo {author} {\bibfnamefont {Anton}\
  \bibnamefont {Zeilinger}},\ }\bibfield  {title} {\enquote {\bibinfo {title}
  {Multiparticle interferometry and the superposition principle},}\ }\href
  {\doibase 10.1063/1.881360} {\bibfield  {journal} {\bibinfo  {journal}
  {Physics Today}\ }\textbf {\bibinfo {volume} {46}},\ \bibinfo {pages}
  {22--29} (\bibinfo {year} {1993})}\BibitemShut {NoStop}%
\bibitem [{\citenamefont {Reck}\ \emph {et~al.}(1994)\citenamefont {Reck},
  \citenamefont {Zeilinger}, \citenamefont {Bernstein},\ and\ \citenamefont
  {Bertani}}]{rzbb}%
  \BibitemOpen
  \bibfield  {author} {\bibinfo {author} {\bibfnamefont {Michael}\ \bibnamefont
  {Reck}}, \bibinfo {author} {\bibfnamefont {Anton}\ \bibnamefont {Zeilinger}},
  \bibinfo {author} {\bibfnamefont {Herbert~J.}\ \bibnamefont {Bernstein}}, \
  and\ \bibinfo {author} {\bibfnamefont {Philip}\ \bibnamefont {Bertani}},\
  }\bibfield  {title} {\enquote {\bibinfo {title} {Experimental realization of
  any discrete unitary operator},}\ }\href {\doibase 10.1103/PhysRevLett.73.58}
  {\bibfield  {journal} {\bibinfo  {journal} {Physical Review Letters}\
  }\textbf {\bibinfo {volume} {73}},\ \bibinfo {pages} {58--61} (\bibinfo
  {year} {1994})}\BibitemShut {NoStop}%
\bibitem [{\citenamefont {Murnaghan}(1962)}]{murnaghan}%
  \BibitemOpen
  \bibfield  {author} {\bibinfo {author} {\bibfnamefont {Francis~D.}\
  \bibnamefont {Murnaghan}},\ }\href@noop {} {\emph {\bibinfo {title} {The
  Unitary and Rotation Groups}}},\ \bibinfo {series} {Lectures on Applied
  Mathematics}, Vol.~\bibinfo {volume} {3}\ (\bibinfo  {publisher} {Spartan
  Books},\ \bibinfo {address} {Washington, D.C.},\ \bibinfo {year}
  {1962})\BibitemShut {NoStop}%
\bibitem [{\citenamefont {Zukowski}\ \emph {et~al.}(1997)\citenamefont
  {Zukowski}, \citenamefont {Zeilinger},\ and\ \citenamefont
  {Horne}}]{zukowski-97}%
  \BibitemOpen
  \bibfield  {author} {\bibinfo {author} {\bibfnamefont {Marek}\ \bibnamefont
  {Zukowski}}, \bibinfo {author} {\bibfnamefont {Anton}\ \bibnamefont
  {Zeilinger}}, \ and\ \bibinfo {author} {\bibfnamefont {Michael~A.}\
  \bibnamefont {Horne}},\ }\bibfield  {title} {\enquote {\bibinfo {title}
  {Realizable higher-dimensional two-particle entanglements via multiport beam
  splitters},}\ }\href {\doibase 10.1103/PhysRevA.55.2564} {\bibfield
  {journal} {\bibinfo  {journal} {Physical Review A}\ }\textbf {\bibinfo
  {volume} {55}},\ \bibinfo {pages} {2564--2579} (\bibinfo {year}
  {1997})}\BibitemShut {NoStop}%
\bibitem [{\citenamefont {Filipp}\ and\ \citenamefont
  {Svozil}(2004)}]{filipp-svo-04-qpoly-prl}%
  \BibitemOpen
  \bibfield  {author} {\bibinfo {author} {\bibfnamefont {Stefan}\ \bibnamefont
  {Filipp}}\ and\ \bibinfo {author} {\bibfnamefont {Karl}\ \bibnamefont
  {Svozil}},\ }\bibfield  {title} {\enquote {\bibinfo {title} {Generalizing
  {T}sirelson's bound on {B}ell inequalities using a min-max principle},}\
  }\href {\doibase 10.1103/PhysRevLett.93.130407} {\bibfield  {journal}
  {\bibinfo  {journal} {Physical Review Letters}\ }\textbf {\bibinfo {volume}
  {93}},\ \bibinfo {pages} {130407} (\bibinfo {year} {2004})},\ \Eprint
  {http://arxiv.org/abs/arXiv:quant-ph/0403175} {arXiv:quant-ph/0403175}
  \BibitemShut {NoStop}%
\bibitem [{\citenamefont {Krenn}\ \emph {et~al.}(2017)\citenamefont {Krenn},
  \citenamefont {Hochrainer}, \citenamefont {Lahiri},\ and\ \citenamefont
  {Zeilinger}}]{PhysRevLett.118.080401}%
  \BibitemOpen
  \bibfield  {author} {\bibinfo {author} {\bibfnamefont {Mario}\ \bibnamefont
  {Krenn}}, \bibinfo {author} {\bibfnamefont {Armin}\ \bibnamefont
  {Hochrainer}}, \bibinfo {author} {\bibfnamefont {Mayukh}\ \bibnamefont
  {Lahiri}}, \ and\ \bibinfo {author} {\bibfnamefont {Anton}\ \bibnamefont
  {Zeilinger}},\ }\bibfield  {title} {\enquote {\bibinfo {title} {Entanglement
  by path identity},}\ }\href {\doibase 10.1103/PhysRevLett.118.080401}
  {\bibfield  {journal} {\bibinfo  {journal} {Physical Review Letter}\ }\textbf
  {\bibinfo {volume} {118}},\ \bibinfo {pages} {080401} (\bibinfo {year}
  {2017})},\ \Eprint {http://arxiv.org/abs/arXiv:1610.00642} {arXiv:1610.00642}
  \BibitemShut {NoStop}%
\bibitem [{\citenamefont {Zou}\ \emph {et~al.}(1991)\citenamefont {Zou},
  \citenamefont {Wang},\ and\ \citenamefont {Mandel}}]{zou-wang-mandel:91a}%
  \BibitemOpen
  \bibfield  {author} {\bibinfo {author} {\bibfnamefont {X.~Y.}\ \bibnamefont
  {Zou}}, \bibinfo {author} {\bibfnamefont {L.~J.}\ \bibnamefont {Wang}}, \
  and\ \bibinfo {author} {\bibfnamefont {L.}~\bibnamefont {Mandel}},\
  }\bibfield  {title} {\enquote {\bibinfo {title} {Induced coherence and
  indistinguishability in optical interference},}\ }\href {\doibase
  10.1103/PhysRevLett.67.318} {\bibfield  {journal} {\bibinfo  {journal}
  {Physical Review Letters}\ }\textbf {\bibinfo {volume} {67}},\ \bibinfo
  {pages} {318--321} (\bibinfo {year} {1991})}\BibitemShut {NoStop}%
\bibitem [{\citenamefont {Wang}\ \emph {et~al.}(1991)\citenamefont {Wang},
  \citenamefont {Zou},\ and\ \citenamefont {Mandel}}]{zou-wang-mandel:91b}%
  \BibitemOpen
  \bibfield  {author} {\bibinfo {author} {\bibfnamefont {L.~J.}\ \bibnamefont
  {Wang}}, \bibinfo {author} {\bibfnamefont {X.~Y.}\ \bibnamefont {Zou}}, \
  and\ \bibinfo {author} {\bibfnamefont {L.}~\bibnamefont {Mandel}},\
  }\bibfield  {title} {\enquote {\bibinfo {title} {Induced coherence without
  induced emission},}\ }\href {\doibase 10.1103/PhysRevA.44.4614} {\bibfield
  {journal} {\bibinfo  {journal} {Physical Review A}\ }\textbf {\bibinfo
  {volume} {44}},\ \bibinfo {pages} {4614--4622} (\bibinfo {year}
  {1991})}\BibitemShut {NoStop}%
\bibitem [{\citenamefont {Ou}(2007)}]{ou-2007}%
  \BibitemOpen
  \bibfield  {author} {\bibinfo {author} {\bibfnamefont {Zhe-Yu~Jeff}\
  \bibnamefont {Ou}},\ }\href {\doibase 10.1007/978-0-387-25554-5} {\emph
  {\bibinfo {title} {Multi-Photon Quantum Interference}}}\ (\bibinfo
  {publisher} {Springer-Verlag US},\ \bibinfo {address} {New York, NY},\
  \bibinfo {year} {2007})\BibitemShut {NoStop}%
\bibitem [{\citenamefont {IBM}(2016)}]{Bennett-IBM-03.05.2016}%
  \BibitemOpen
  \bibfield  {author} {\bibinfo {author} {\bibnamefont {IBM}},\ }\href
  {https://www.youtube.com/watch?v=9q-qoeqVVD0} {\enquote {\bibinfo {title}
  {{C}harles {B}ennett -- a founder of quantum information theory},}\ }
  (\bibinfo {year} {2016}),\ \bibinfo {note} {may 3rd, 2016, accessed July
  16th, 2016}\BibitemShut {NoStop}%
\bibitem [{\citenamefont
  {Schr{\"{o}}dinger}(1935{\natexlab{a}})}]{schrodinger}%
  \BibitemOpen
  \bibfield  {author} {\bibinfo {author} {\bibfnamefont {Erwin}\ \bibnamefont
  {Schr{\"{o}}dinger}},\ }\bibfield  {title} {\enquote {\bibinfo {title} {Die
  gegenw{\"{a}}rtige {S}ituation in der {Q}uantenmechanik},}\ }\href {\doibase
  10.1007/BF01491891, 10.1007/BF01491914, 10.1007/BF01491987} {\bibfield
  {journal} {\bibinfo  {journal} {Naturwissenschaften}\ }\textbf {\bibinfo
  {volume} {23}},\ \bibinfo {pages} {807--812, 823--828, 844--849} (\bibinfo
  {year} {1935}{\natexlab{a}})}\BibitemShut {NoStop}%
\bibitem [{\citenamefont {Trimmer}(1980)}]{schrodinger-en-10.2307/986572}%
  \BibitemOpen
  \bibfield  {author} {\bibinfo {author} {\bibfnamefont {John~D.}\ \bibnamefont
  {Trimmer}},\ }\bibfield  {title} {\enquote {\bibinfo {title} {The present
  situation in quantum mechanics: A translation of {S}chr\"odinger's ``cat
  paradox'' paper},}\ }\href {http://www.jstor.org/stable/986572} {\bibfield
  {journal} {\bibinfo  {journal} {Proceedings of the American Philosophical
  Society}\ }\textbf {\bibinfo {volume} {124}},\ \bibinfo {pages} {323--338}
  (\bibinfo {year} {1980})}\BibitemShut {NoStop}%
\bibitem [{\citenamefont
  {Schr{\"{o}}dinger}(1935{\natexlab{b}})}]{CambridgeJournals:1737068}%
  \BibitemOpen
  \bibfield  {author} {\bibinfo {author} {\bibfnamefont {Erwin}\ \bibnamefont
  {Schr{\"{o}}dinger}},\ }\bibfield  {title} {\enquote {\bibinfo {title}
  {Discussion of probability relations between separated systems},}\ }\href
  {\doibase 10.1017/S0305004100013554} {\bibfield  {journal} {\bibinfo
  {journal} {Mathematical Proceedings of the Cambridge Philosophical Society}\
  }\textbf {\bibinfo {volume} {31}},\ \bibinfo {pages} {555--563} (\bibinfo
  {year} {1935}{\natexlab{b}})}\BibitemShut {NoStop}%
\bibitem [{\citenamefont
  {Schr{\"{o}}dinger}(1936)}]{CambridgeJournals:2027212}%
  \BibitemOpen
  \bibfield  {author} {\bibinfo {author} {\bibfnamefont {Erwin}\ \bibnamefont
  {Schr{\"{o}}dinger}},\ }\bibfield  {title} {\enquote {\bibinfo {title}
  {Probability relations between separated systems},}\ }\href {\doibase
  10.1017/S0305004100019137} {\bibfield  {journal} {\bibinfo  {journal}
  {Mathematical Proceedings of the Cambridge Philosophical Society}\ }\textbf
  {\bibinfo {volume} {32}},\ \bibinfo {pages} {446--452} (\bibinfo {year}
  {1936})}\BibitemShut {NoStop}%
\bibitem [{\citenamefont {Wootters}(1990)}]{wootters-1990-localaccOQStates}%
  \BibitemOpen
  \bibfield  {author} {\bibinfo {author} {\bibfnamefont {William~K.}\
  \bibnamefont {Wootters}},\ }\bibfield  {title} {\enquote {\bibinfo {title}
  {Local accessibility of quantum states},}\ }in\ \href@noop {} {\emph
  {\bibinfo {booktitle} {Complexity, Entropy, and the Physics of
  Information}}},\ \bibinfo {series and number} {{SFI} Studies in the Sciences
  of Complexity, Vol. {VIII}},\ \bibinfo {editor} {edited by\ \bibinfo {editor}
  {\bibfnamefont {Wojciech~Hubert}\ \bibnamefont {Zurek}}}\ (\bibinfo
  {publisher} {Addison-Wesley},\ \bibinfo {address} {Boston},\ \bibinfo {year}
  {1990})\ pp.\ \bibinfo {pages} {39--46}\BibitemShut {NoStop}%
\bibitem [{\citenamefont {Mermin}(1998)}]{mermin:753}%
  \BibitemOpen
  \bibfield  {author} {\bibinfo {author} {\bibfnamefont {David~N.}\
  \bibnamefont {Mermin}},\ }\bibfield  {title} {\enquote {\bibinfo {title}
  {What is quantum mechanics trying to tell us?}}\ }\href {\doibase
  10.1119/1.18955} {\bibfield  {journal} {\bibinfo  {journal} {American Journal
  of Physics}\ }\textbf {\bibinfo {volume} {66}},\ \bibinfo {pages} {753--767}
  (\bibinfo {year} {1998})},\ \Eprint
  {http://arxiv.org/abs/arXiv:quant-ph/9801057} {arXiv:quant-ph/9801057}
  \BibitemShut {NoStop}%
\bibitem [{\citenamefont {Zeilinger}(1997)}]{Zeilinger-97}%
  \BibitemOpen
  \bibfield  {author} {\bibinfo {author} {\bibfnamefont {Anton}\ \bibnamefont
  {Zeilinger}},\ }\bibfield  {title} {\enquote {\bibinfo {title} {Quantum
  teleportation and the non-locality of information},}\ }\href {\doibase
  10.1098/rsta.1997.0138} {\bibfield  {journal} {\bibinfo  {journal}
  {Philosophical Transactions of the Royal Society of London A}\ }\textbf
  {\bibinfo {volume} {355}},\ \bibinfo {pages} {2401--2404} (\bibinfo {year}
  {1997})}\BibitemShut {NoStop}%
\bibitem [{\citenamefont {Zeilinger}(1999)}]{zeil-99}%
  \BibitemOpen
  \bibfield  {author} {\bibinfo {author} {\bibfnamefont {Anton}\ \bibnamefont
  {Zeilinger}},\ }\bibfield  {title} {\enquote {\bibinfo {title} {A
  foundational principle for quantum mechanics},}\ }\href {\doibase
  10.1023/A:1018820410908} {\bibfield  {journal} {\bibinfo  {journal}
  {Foundations of Physics}\ }\textbf {\bibinfo {volume} {29}},\ \bibinfo
  {pages} {631--643} (\bibinfo {year} {1999})}\BibitemShut {NoStop}%
\bibitem [{\citenamefont {Svozil}(2017)}]{svozil-2016-pu-book}%
  \BibitemOpen
  \bibfield  {author} {\bibinfo {author} {\bibfnamefont {Karl}\ \bibnamefont
  {Svozil}},\ }\href@noop {} {\emph {\bibinfo {title} {Physical [A]Causality.
  {D}eterminism, Randomness and Uncaused Events}}}\ (\bibinfo  {publisher}
  {Springer},\ \bibinfo {address} {Berlin, Heidelberg, New York},\ \bibinfo
  {year} {2017})\ \bibinfo {note} {in print}\BibitemShut {NoStop}%
\bibitem [{\citenamefont {Mermin}(2007)}]{mermin-07}%
  \BibitemOpen
  \bibfield  {author} {\bibinfo {author} {\bibfnamefont {David~N.}\
  \bibnamefont {Mermin}},\ }\href
  {http://people.ccmr.cornell.edu/~mermin/qcomp/CS483.html} {\emph {\bibinfo
  {title} {Quantum Computer Science}}}\ (\bibinfo  {publisher} {Cambridge
  University Press},\ \bibinfo {address} {Cambridge},\ \bibinfo {year}
  {2007})\BibitemShut {NoStop}%
\bibitem [{\citenamefont {Schwinger}(1960)}]{Schwinger.60}%
  \BibitemOpen
  \bibfield  {author} {\bibinfo {author} {\bibfnamefont {Julian}\ \bibnamefont
  {Schwinger}},\ }\bibfield  {title} {\enquote {\bibinfo {title} {Unitary
  operators bases},}\ }\href {\doibase 10.1073/pnas.46.4.570} {\bibfield
  {journal} {\bibinfo  {journal} {Proceedings of the National Academy of
  Sciences (PNAS)}\ }\textbf {\bibinfo {volume} {46}},\ \bibinfo {pages}
  {570--579} (\bibinfo {year} {1960})}\BibitemShut {NoStop}%
\bibitem [{\citenamefont {Macchiavello}\ \emph {et~al.}(2001)\citenamefont
  {Macchiavello}, \citenamefont {Palma},\ and\ \citenamefont
  {Zeilinger}}]{Macchiavello-2001}%
  \BibitemOpen
  \bibfield  {author} {\bibinfo {author} {\bibfnamefont {C.}~\bibnamefont
  {Macchiavello}}, \bibinfo {author} {\bibfnamefont {G.~M.}\ \bibnamefont
  {Palma}}, \ and\ \bibinfo {author} {\bibfnamefont {Anton}\ \bibnamefont
  {Zeilinger}},\ }\href {\doibase 10.1142/9789810248185} {\emph {\bibinfo
  {title} {Quantum Computation and Quantum Information Theory}}}\ (\bibinfo
  {publisher} {World Scientific},\ \bibinfo {address} {Singapore},\ \bibinfo
  {year} {2001})\BibitemShut {NoStop}%
\end{thebibliography}


\end{document}